\documentclass[a4paper,referee]{raa}           
\usepackage{graphicx,times}
\usepackage{natbib}
\usepackage{amssymb,amsmath}
\bibpunct{(}{)}{;}{a}{}{,}

\usepackage[a4paper=true,pagebackref=true]{hyperref}
\hypersetup{pdftitle = The title of my PDF, pdfauthor = My name, pdfsubject= The subject, pdfkeywords = keyword1 keyword2 keyword3} 
\hypersetup{colorlinks = true, linkcolor = green, anchorcolor = red, citecolor = blue, filecolor = red, pagecolor = red, urlcolor = red}

\begin{document}

   \title{The Mini-SiTian Array: first-two-year operation}

 \volnopage{ {\bf 20XX} Vol.\ {\bf X} No. {\bf XX}, 000--000}
   \setcounter{page}{1}

   \author{Min He\inst{1}, Hong Wu\inst{1,3}, Liang Ge\inst{1,4}, Jian-feng Tian\inst{1}, Zheng Wang\inst{2}, Hai-yang Mu\inst{1}, Yu Zhang\inst{1}, Yang Huang\inst{3,1}, Jie Zheng\inst{1}, Zhou Fan\inst{1,3}, Zheng-yang Li\inst{5}, Hong-rui Gu\inst{3,1}, Heng-geng Han\inst{1,3}, Kai Xiao\inst{3,1}, Zhi-rui Li\inst{3,1}, Jun-jie Jin\inst{1}, Bei-chuan Wang\inst{3,1}, Jun Ma\inst{3,1}, Jin-hang Zou\inst{2}, Ying Wu\inst{1}, Jiu-peng Guo\inst{1}, Li-guo Fang\inst{1}, Zhi-gang Hou\inst{1}, Bo-wen Zhang\inst{3,1}, Yun-fei Xu\inst{2,3,6}, Yi-ming Mao\inst{3,1}, Shuai Liu\inst{1,3,7}, Fang-zhou Ren\inst{3,1}, Cun-shi Wang\inst{3,1}, Xue Li\inst{3,1}, Yong-xin Wu\inst{3,1}, Chuan-jie Zheng\inst{3,1}, Yi-yang Lin\inst{3,1}, Shun-xuan He\inst{3,1}, Kun Xu\inst{8}, Yi-nan Zhu\inst{1}, Zhi-jun Tu\inst{3,1}, Xin-lin Zhao\inst{3,1}, Yong-kang Sun\inst{3,1}, Hua Bao\inst{9,10}, Xue-ang Sun\inst{1}, Ying-zhen Cui\inst{3,1}, Lan-ya Mou\inst{3,1}, Rui-feng Shi\inst{3,1}, Jing-hang Shi\inst{3,1}, Xun-hao Chen\inst{3,1}, Yu-liang Hong\inst{3,1}, Hao-miao Huang\inst{3,1}, Zi-kun Lin\inst{3,1}, Ze-yang Pan\inst{3,1}, Rui-ning Zhao\inst{3,1}, Ying-jie Cai\inst{3,1}, Rui Wang\inst{1}, Jun-bo Zhang\inst{1}, Jin-hu Wang\inst{1}, Wan Zhou\inst{1}, Sheng-ming Li\inst{1}, Jing Ren\inst{3,1}, Xiao-han Chen\inst{3,1}, Xiang-yu Li\inst{3,1}, Li-yue Zhang\inst{3,1}, Hao Mo\inst{3,1}, Yu-yang Zhou\inst{3,1}, Jia-hui Wang\inst{3,1}, Bu-hui Lv\inst{3,1}, Zhao-hui Shang\inst{2}, Yun-ning Zhao\inst{3,1}, Jin-lei Zhang\inst{1}, Yue Sun\inst{1}, Xiao-peng Liu\inst{11}, Zi-jian Han\inst{5}, Yong-na Mao\inst{3,1}, Ji-feng Liu\inst{1,3}}

   \institute{Key Laboratory of Optical Astronomy, National Astronomical Observatories, Chinese Academy of Sciences, 20A Datun Road, Chaoyang District, Beijing 100101, People’s Republic of China; {\it hemin@nao.cas.cn}\\
        \and National Astronomical Observatories, Chinese Academy of Sciences, Beijing 100101, People’s Republic of China\\
        \and School of Astronomy and Space Science, University of Chinese Academy of Sciences, 19A Yuquan Road, Shijingshan District, Beijing 100049, People’s Republic of China\\
        \and Institute for Frontiers in Astronomy and Astrophysics, Beijing Normal University, Beijing 100875, China\\
        \and Nanjing Institute of Astronomical Optics \& Technology, Chinese Academy of Sciences, Nanjing 211135, Jiangsu, People’s Republic of China\\
        \and National Astronomical Data Center, Beijing 100101, People’s Republic of China\\
        \and Instituto de Astrof\'{i}sica de Canarias, V\'{i}a L\'{a}ctea, 38205 La Laguna, Tenerife, Spain\\
        \and School of Science, Qingdao University of Technology, Qingdao 266525, People’s Republic of China\\
        \and College of Science, Tibet University, Lhasa 850000, People’s Republic of China\\
        \and The Key Laboratory of Cosmic Rays (Tibet University), Ministry of Education, Lhasa 850000, People’s Republic of China;
        \and Department of Physics, Hebei Normal University, Shijiazhuang 050024, , People’s Republic of China\\
\vs \no
   {\small Received 20xx December XX; accepted 202X XX XX}
}

\abstract{The SiTian project, designed to utilize 60 telescopes distributed across multiple sites in China, is a next-generation  time-domain survey initiative. As a pathfinder for the SiTian project, the Mini-SiTian (MST) has been proposed and implemented to test the SiTian's brain and data pipeline, and to evaluate the feasibility of its technology and science cases. Mounted at the Xinglong Observatory, the MST project comprises three 30 cm telescopes and has been operated since Nov. 2022. Each telescope of the MST possesses a large field of view, covering $2.29\dg \times 1.53\dg$ FOV, and is equipped with $g'$, $r'$ and $i'$ filters, respectively. Acting as the pioneer of the forthcoming SiTian project, the MST is dedicated to the discovery of variable stars, transients, and outburst events, and has already obtained some interesting scientific results. In this paper, we will summarize the first-two-year operation of the MST project. 
}

   \authorrunning{Min He et al. }            
   \titlerunning{First-two-year Operation of the Mini-SiTian}  
   \maketitle

%
\section{Introduction}          
\label{sect:intro}

SiTian (\citealt{LiuJF+etal+2020}) is a next generation time-domain survey project. It is designed to use 60 telescopes, which located at several sites in China, to scan a sky field survey of ~10,000 square degrees using three bands in each half-hour. Based on such high survey efficiency, this project will be able to discover and identify transient variable sources in real time, triggering follow-up observations with spectroscopic telescopes or collaborating with space telescopes, such as the Chinese Enstein probe (\citealt{2022hxga.book...86Y}) and SVOM (\citealt{2022IJMPD..3130008A}).

The entire SiTian project comprises the SiTian brain, telescope arrays and data storage system. The SiTian brain serves as a platform to coordinate the operation of telescopes, the real-time processing of data, the transmission of data, and the triggering of alerts for transients. As a precursor to the SiTian project, the Mini-SiTian project (hereafter MST) has been proposed and implemented to test the SiTian brain, data pipeline, instrumental technology and early science cases. 

The MST consists of three 30 cm diameter telescopes, which can simulate a single node of the SiTian array. The desired capabilities of the MST include:
(a) the ability to complete observations automatically,
(b) the capability to acquire raw data and conduct processing in real time (less than 5 minutes),
(c) the capacity for fast communication with larger telescopes for follow-up observations,
(d) the flexibility to accommodate various programs with different observational constraints, and
(e) the implementation of fully automated operations through robotic or AI-driven systems.
Additionally, the MST is anticipated to play a significant role in the education of observational astronomy by providing numerous opportunities for young graduate students to develop skills in conducting observations and analyzing astronomical data.

The MST has been in operation for two years since November 2022. In this paper, we will summarize the pilot operation of the MST project from November 2022 to June 2024. An brief overview of the MST project will be presented in Section~\ref{sect:overview}. Section~\ref{sect:strategy} will describe the observing strategies, including target selection and the preparation of calibration data. Daily observations and some data results will be described in Sections~\ref{sect:operation} and~\ref{sect:performance}. Finally, in Section~\ref{sect:sum}, we will provide a summary of the MST project and discuss the next steps for this project.

\section{The MST Overview}
\label{sect:overview}

\begin{figure}[!t]
  \centering
  \subfloat{
    \includegraphics[width=\textwidth]{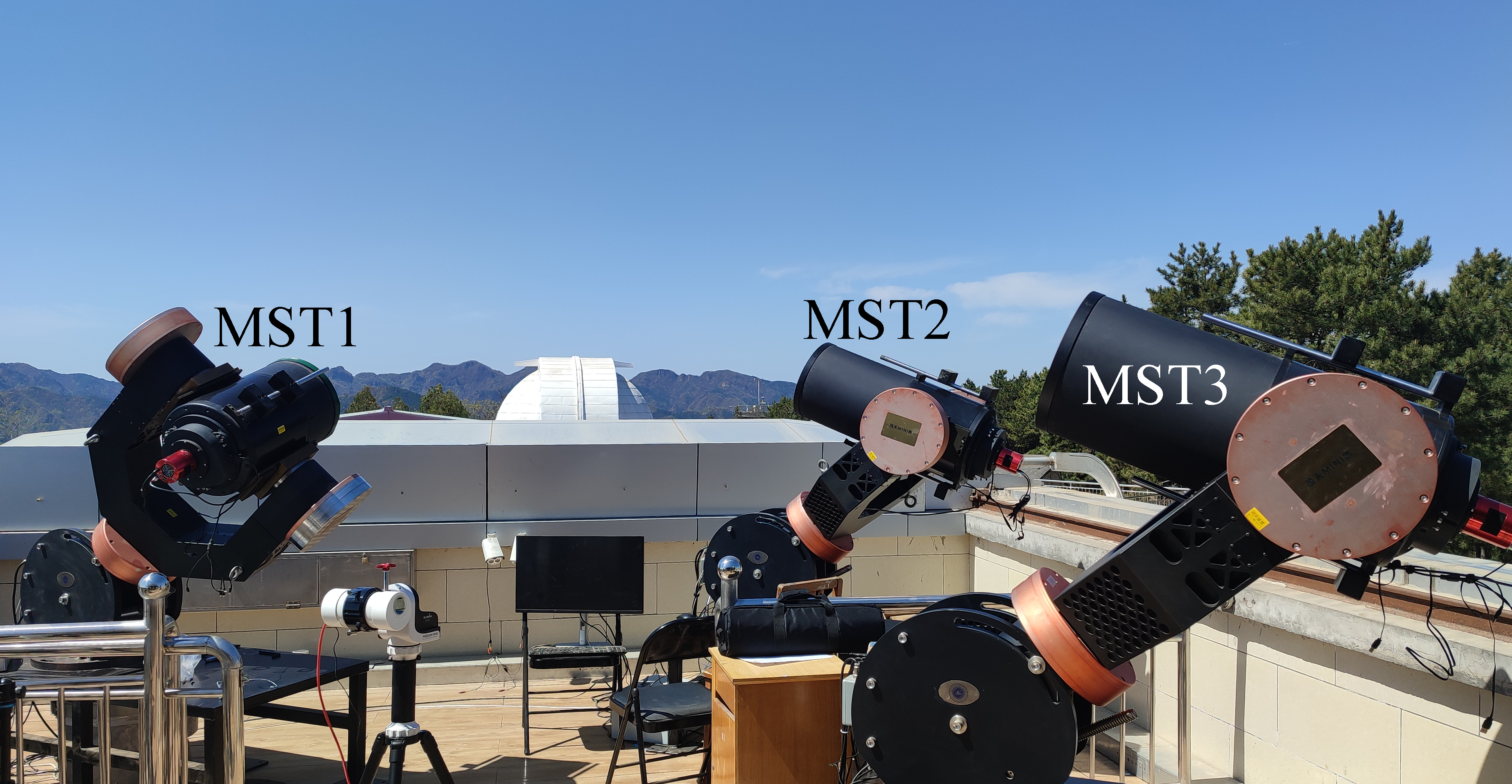}}\\
   \caption{Overview of the MST telescope array.} 
   \label{mini}
\end{figure}

The MST is composed of three 30 cm catadioptric Schmidt telescopes: mini-001 (hereafter MST1), mini-002 (hereafter MST2) and mini-003 (hereafter MST3). These telescopes share a similar optical design with the CSTAR telescope and are referred to as the MEGA-CSTAR optic system (\citealt{hanzi+et+al+2025}) (see Figure~\ref{mini}). They are manufactured by Astronomical Instruments Company Limited (AIC) and each is equipped with an equatorial mount, a $5\dg $ diameter field of view (FOV), and a focal ratio of f/3. The 30 cm catadioptric Schmidt telescope consists of the corrector lens group 1 and the corrector lens group 2 (see Figure 1 of \cite{hanzi+et+al+2025}). For a detailed discussion of the optical design of this telescope, please refer to \cite{hanzi+et+al+2025} in this volume. 

\begin{table}
\begin{center}
\caption[]{Summary of the Mini-SiTian telescope array (MST)}\label{Tab1}
 \begin{tabular}{l}
  \hline\noalign{\smallskip}
Telescope:\\
- Primary mirror: 30 cm in diameter for all three telescopes\\
- Optics: Catadioptric Schimidt\\
- Field of View (FOV): $5\dg \times 5\dg$\\
- Focal ratio: f/3 \\
- Mount: Fork-type equatorial\\
- Installed site: Xinglong Observatory\\
  \hline\noalign{\smallskip}
Camera:\\
- $9,576\times 6,388$ ZWO ASI6200MM Pro CMOS\\
- $3.76\mu $m pixel size, 0.862 arcsec per pixel, $2.29\dg \times 1.53\dg$ FOV\\
- Quantum efficiency of $~90\%$ in $g'$-band, $~70\%$ in $r'$-band and $~30\%$ in $i'$-band\\
- CMOS cooled down to 30 degree Celsius lower than environment temperature\\
  \hline\noalign{\smallskip}
Filter and focuser:\\
- One filter for each telescope, $i'$, $g'$ and $r'$ band for MST1, MST2 and MST3, respectively\\
- Self-made stepper motor-driven focuser by AIC\\
  \noalign{\smallskip}\hline
\end{tabular}
\end{center}
\end{table}

\begin{figure}[!t]
  \centering
  \includegraphics[width=\textwidth, angle=0]{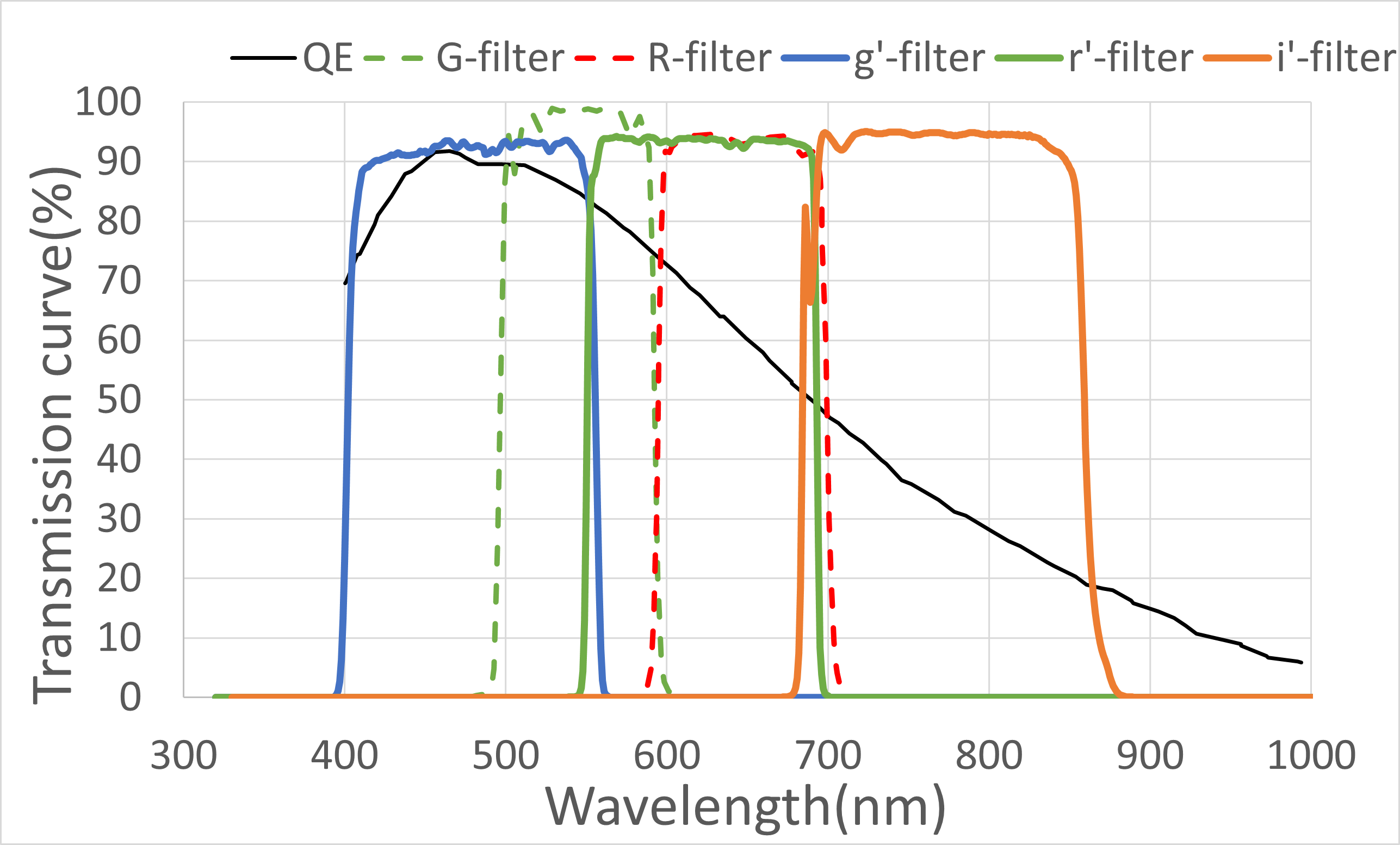}
  \caption{The black curve represents the quantum efficiency (QE) curve of ZWO ASI6200MM Pro CMOS camera, which shows higher sensitivity in the blue band. The colored curves represent the transmission curves of filters of the MST. The dotted curves correspond the filters used before February 7, 2023, which were the $G$ and $R$-band filters, and the solid curves correspond to the filters used after February 7, 2023, which are Solan-like $g'$, $r'$, and $i'$-band filters.} 
  \label{qe}
\end{figure}

The terminal instruments include a ZWO ASI6200MM Pro CMOS camera, one filter, and a stepper motor-driven focuser by the AIC. The size of ZWO ASI6200MM Pro CMOS camera is $9,576\times 6,388$ pixels, the pixel size is $3.76\mu $m and the pixel scale is $0.862''$/pixel, which lead to a camera FOV of $2.29\dg \times 1.53\dg$. The quantum efficiency of the camera is higher in the bluer range. The filters are similar to the SDSS's system filters $i'$, $g'$, and $r'$, and are mounted in MST1, MST2, and MST3, respectively (see Figure~\ref{qe}) since February 7, 2023. Before this date, we use $G$ and $R$-band filters to replace the $g'$, and $r'$-band filters in MST2 and MST3. Figure~\ref{qe} also shows the quantum efficiency (hereafter QE) curve of the ZWO ASI6200MM Pro CMOS camera. All these parameters of instruments are summarized in Table~\ref{Tab1}.

\begin{figure}
   \centering
   \includegraphics[width=\textwidth, angle=0]{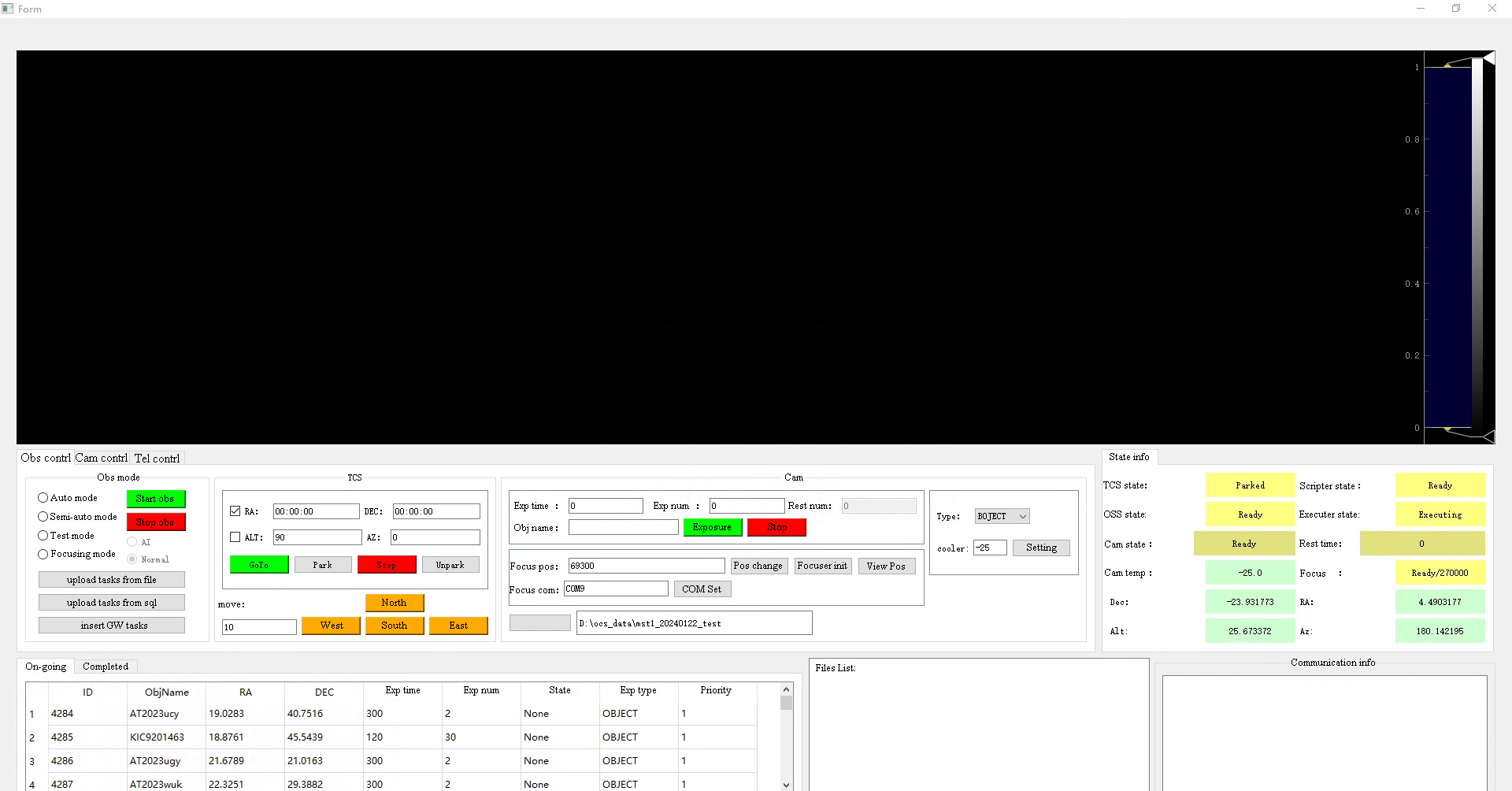}
   \caption{The observational control system of the MST, which is developed in-house based on the python environment.} 
   \label{ocs}
\end{figure}

The observational control system (hereafter OCS), developed in-house based on the python environment (\citealt{wangzheng+et+al+2025}), is shown in Figure~\ref{ocs}. This system integrates the functions of telescope control, camera control, observing plan control, and image display, and it offers four operational modes for different situations: automatic mode, semi-automatic mode, test mode, and focusing mode. The automatic mode is designed for executing automatic observing plans, and can determine whether the next target is within the observing field; if not, it skips the target. In the semi-automatic mode, the OCS observes only one selected target per action. The test mode is used for debugging actions, such as slewing and exposure tests. The focusing mode is intend for focusing actions with fast readout, outputting only the central region ($1,912 \times 1,272$ pixels) of a full image.

\section{Observing strategy}
\label{sect:strategy}
As a pathfinder for the SiTian project, the MST is primarily used to discover new variables and to follow-up some transient events, such as gravitational waves (GWs), comets, supernovae and tidal disruption events (TDEs) (\citealt{han+et+al+2025}) under continuous exposures.

\subsection{Calibration images}
Calibration images are also required for daily observation, including biases and flats. All these images are taken before scientific observations commence. Bias frames are captured with the telescope and dome fully covered to ensure no light enters, as the CMOS camera lacks a mechanical shutter. To achieve uniform illumination for sky flats with a large FOV telescope, a flat plate is placed over the telescope tube to diffuse the skylight. However, using the flat plate significantly reduces the light intensity. To compensate, we take flat-field exposures before sunset, maintaining ADU values between 15,000 and 50,000 to ensure a high enough signal-to-noise ratio for the corrections of pixel-to-pixel variations.

In addition to the sky flats, we also take some supersky flats during the new moon phase. These supersky flats are taken only during clear weather and the fields to be observed for them are positioned near the zenith of the sky.

\subsection{General survey}
For general surveys, our strategy involves selecting specific fields and conducting long-term monitoring to discover new variables and transients. During the first year of the MST operation, we tested its detection capabilities by selecting three test sky fields, designated as f01, f02 and f03. 

\begin{figure}
   \centering
   \includegraphics[width=\textwidth, angle=0]{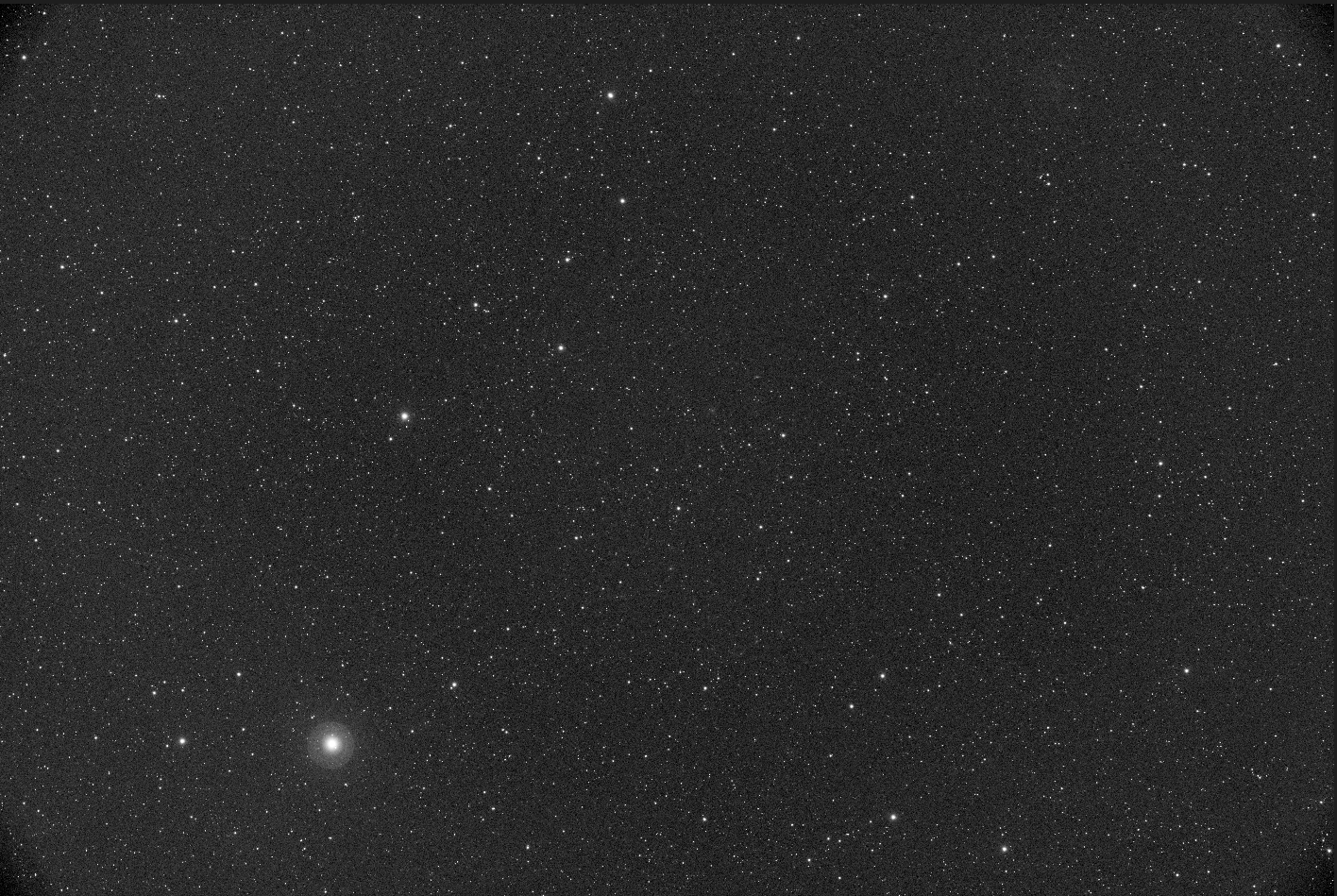}
   \caption{The MST image of the test sky field: f01, which contains one galaxy group, several bright stars, 5 known short-period variables and $\sim 1,000$ LAMOST targets.} 
   \label{f01}
\end{figure}

The f01 field (shown in Figure~\ref{f01}) includes:
	1) {\bf One galaxy group}, which enhances the possibility of detecting supernovae.
	2) {\bf Several bright stars}, useful for assessing the impact of image ghosts.
	3) {\bf Five known short-period variable stars}, which can be utilized to evaluate the image processing system's capability for detecting variability.
	4) {\bf Over 1,000 LAMOST targets} near the Galactic disk plane, suitable for examining variations in FWHM and pixel scale in the image data, as well as analyzing the performance of stellar flat-fielding. 

The f02 and f03 fields were defined using similar criteria as f01, with their R.A. values increased by 1 and 2 hours, respectively, compared to f01. All the three fields have been planned to be exposured throughout the entire nights over a period of one or two months in order to collect sufficient image data.

In the second year of the MST operation, we have selected 87 sky fields for a TDEs \& supernova survey. These 87 sky fields are chosen based on three criteria: (1) the declination of them are higher than $-20\dg$ ; (2) avoiding the region where $|b|<20\dg$ ($b$ is the Galactic latitude) to reduce reddening; (3) selecting fields with a high star formation rate or a large number of galaxies. For daily observations, we plan to allocate 2–3 hours to observe one-third of these fields, which will enable the TDE and supernova survey to cycle every three days.

\subsection{GW follow-up survey}

The GW follow-up survey is based on the O4 project of LIGO \footnote{http://gracedb.ligo.org}. We obtain the information of each GW event published by the alert from the LIGO website as soon as they are available. If the `significant' of event is marked as `true', then we will produce a sky field list, which covers the sky areas observable on our site at that night for the event. The order of sky field list is determined not only by the probability provided by the O4 but also by the turnaround time of telescope to ensure more efficient observation times. The detailed strategy of GW plan can be seen in \cite{2020PASP..132j4501X} and \cite{2024PASP..136h4501M}.

We finally selected the top $200$ deg$^2$ sky fields for observation. Since the primary goal of the GW follow-up survey is to search for optical counterparts of GW events, we do not prioritize their color. The sky field list is divided into three distinct observing plans, one for each MST telescope. These plans are then provided to the daily observing assistants, who upload them to the OCS of the respective telescopes.

Given the highest priority of the GW follow-up survey, upon receiving a GW alert, the daily general observations of the MST will be immediately interrupted to begin the follow-up survey. However, if the MST is unable to observe due to poor weather or other circumstances within 24 hours of receiving the alert, we will cancel the follow-up plan for that alert.

\subsection{Others}
In addition to conducting general surveys and following up on GW events, the MST also observes other transients discovered by different surveys, such as comets, supernovae, and TDEs. 

\subsubsection{Comets}
The study of comets is key to understanding the formation and evolution of matter in the early Solar System (\citealt{1993prpl.conf.1177M,2000prpl.conf.1159I}). And the cometary spectra at special phase are very important for analyzing the composition and the kinematical and spatial structure of comets (\citealt{1948C&T....64...16S,1948AnAp...11..124S,1950ApJ...111..375W,1965ARA&A...3..351A,1986A&A...160L..11D,1986ApJ...310L..55S,1986Natur.324..441W,1986Sci...232.1523M,2011ARA&A..49..471M}). Therefore, the photometric monitoring of comets is very useful for capturing opportunities to observe the spectra of comets during special phases, such as the phase of maximum brightness of cometary nucleus.

\begin{figure}
   \centering
   \includegraphics[width=\textwidth, angle=0]{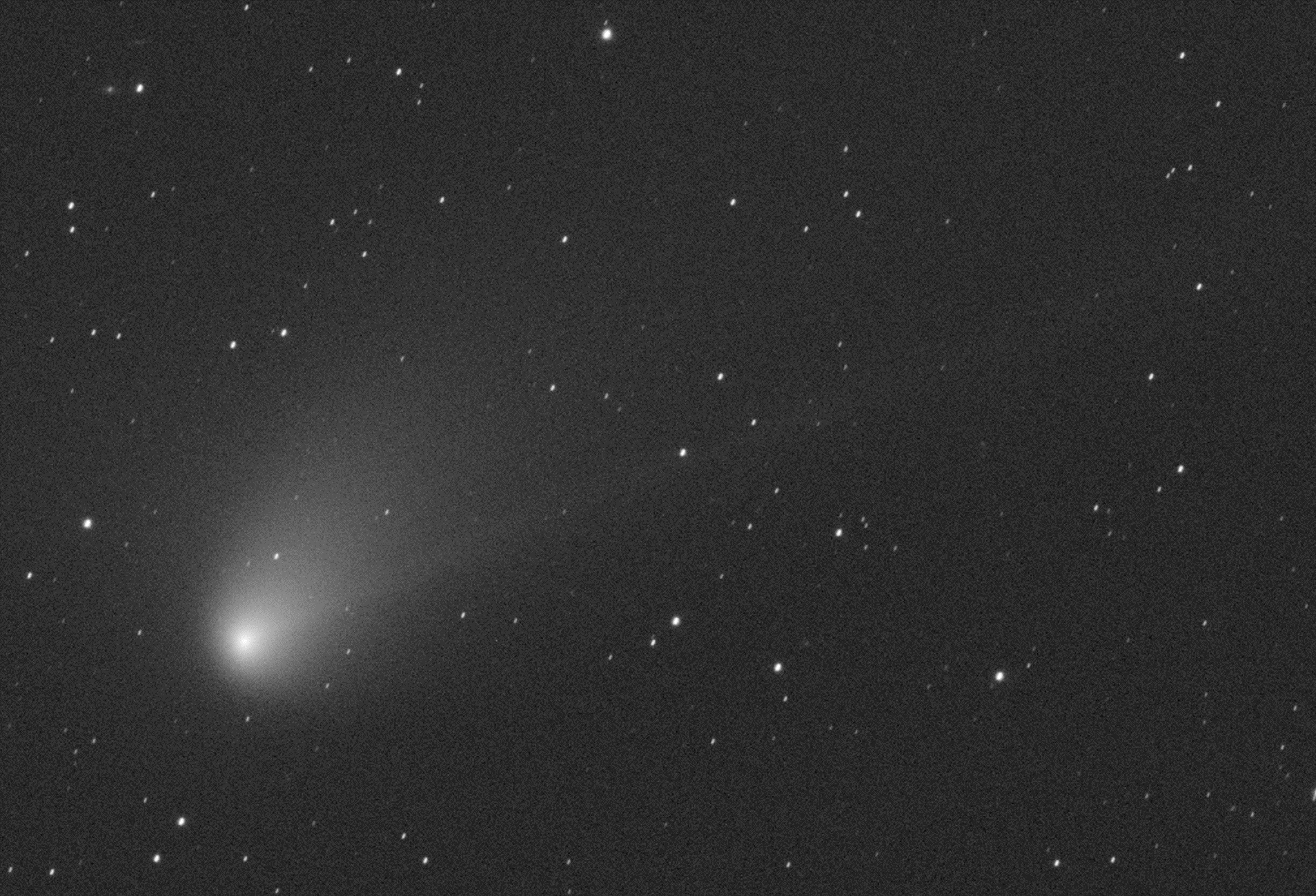}
   \caption{This image of comet C/2022 E3 clearly shows both the comet's tail and core, highlighting the excellent tracking ability and imaging quality of the MST.} 
   \label{comet}
\end{figure}

Since the fast moving speed of comets, the tracking ability and imaging quality of the MST for this kind of sources require to be tested. During the first-two-year operation of the MST, we have monitored two comets: C/2022 E3 and 12P in January 2023 and October 2023, respectively. The MST has observed each comet for nearly a month, and an image of C/2022 E3 is displayed in Figure~\ref{comet}. The structures of the cometary tail and cometary nuclei are clearly presented in this frame. The data of them obtained by the MST has been successfully utilized for follow-up spectroscopic observations at the phase of maximum luminosity by monitoring the light variation of the cometary nuclei.

\subsubsection{Supernovae and TDEs}
We select supernovae and TDE candidates from the Transient Name Server (TNS) website \footnote{\url{www.wis-tns.org}}, where new astronomical transients (ATs) are reported every day. Each AT report represents a potential transient event, with the majority being supernovae and TDEs. We check new AT reports daily and select supernovae and TDE candidates for observation based on three key considerations.

First, we exclude potential transients whose altitude is too low to ensure high data quality. Considering the geographical location of the MST, targets with a declination lower than $-20{\dg}$ are excluded, and the targets with altitudes lower than $20^{\circ}$ throughout the entire night are also excluded. Secondly, most of the targets located far from a galaxy are eliminated to ensure a high possibility to be extra-galactic supernovae. Last but not least, whether the target is valuable to be observed, will also be considered. Since the early light of supernovae provides essential information to constrain its progenitors (\citealt{2010ApJ...708.1025K}), it is crucial to capture the light curves of supernovae as early as possible. We primarily focus on targets with low redshift (mainly focusing on $z < 0.02$) to ensure that we observe supernovae while the light curve of them are still on the early stage of rising. Additionally, if the target was observed by Zwicky Transient Facility (\citealt{2019PASP..131a8002B}), the likelihood of it being a supernova, which provided by the ALeRCE stamp classifier (\citealt{2021AJ....161..242F}), would also be taken into account.

\subsubsection{Stellar flares}
Stellar flares are energetic events caused by the re-connection of stellar magnetic fields, and they are very useful for studying the solar-stellar connections. Among many bands observations of this phenomenon, optic flares typically have high energy and sometimes this energy even higher than $10^{34}$ erg for the superflares in solar-type stars (\citealt{2012Natur.485..478M,2017ApJ...849...36Y}).

In the Kepler and TESS fields, thousands of flaring stars have been identified. We carry out a long-term monitoring of the flaring stars, which are located in Kepler and TESS fields, and hope to help constrain the solar and stellar dynamo theories.

\section{Observing operation}
\label{sect:operation}
\begin{figure}
   \centering
   \includegraphics[width=\textwidth, angle=0]{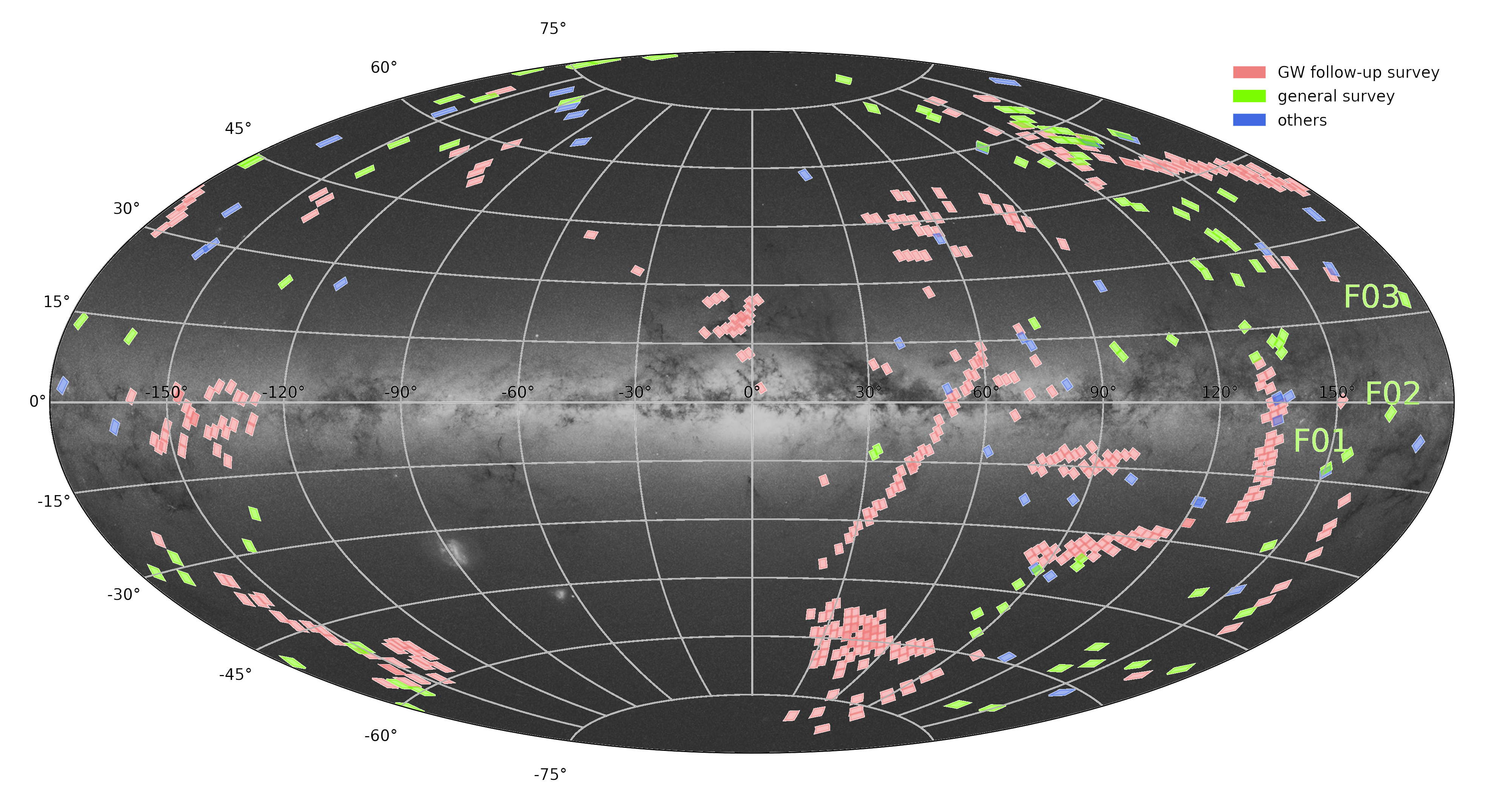}
   \caption{Fields observed during the first-two-year operation of the MST for various scientific purposes, with the background image of our Galaxy taken from Gaia.} 
   \label{fields}
\end{figure}

Two of the three MST telescopes, MST2 and MST3, were mounted in the autumn of 2022. And they have been in operation from November 2022 to June 2024. Then, they had a shutdown in July and Autumn because of the rainy season. The MST1 is scheduled to begin operation from September 2024. During the first-two-year operation, we have observed 92 general survey fields, 1 fields of active stars potentially with stellar flares, and followed up 37 transients from TNS, 33 GWs alerts from LIGO and 2 comets, as depicted in Figure~\ref{fields}. In this section, we will present the pilot operation of the first-two-year, which includes general plan and dynamic plan. 

\subsection{General plan}
General plan of the MST includes long-time observation of special fields. The object lists for each science topic are submitted by the P.I. and compiled by the operation administrator. Subsequently, based on the daily charting curve for each object and the horizon limit of $16{\dg}$, daily observing plan will be produced by the operation administrator and sent to the daily observational assistants. Daily observation begins at $-12{\dg}$ of sunset and ends at $-12{\dg}$ of sunrise. During the first-two-year operation, we have observed 93 fields, including f01, f02, and f03 for 73 days, UH01 and UH02 for 5 months, 1 Kepler fields for 4 months, 87 TDEs \& supernova searching fields for 10 months).

\subsection{Dynamic Plan}
For transient events, we implement a dynamic daily plan that prioritizes tasks based on their urgency. Gravitational waves (GWs) are assigned the highest priority, followed by supernovae and tidal disruption events (TDEs), which are ranked below GWs but above general surveys due to their relatively longer transient durations. If the dynamic observing proposals are submitted during the daytime, we adjust the daily plan to ensure the exposure of high-priority sources. However, if the proposals are submitted at night, which are mostly GWs, we interrupt the on-going exposure to start the observation of GW events and resume the general plan when we have free time again. We have followed up 37 transients from TNS and 2 comets for several days, and also have tested following-up GWs survey for 33 GW alerts from LIGO O4.

\section{Observing Performance}
\label{sect:performance}

\begin{figure}
   \centering
   \subfloat{
     \includegraphics[width=\textwidth]{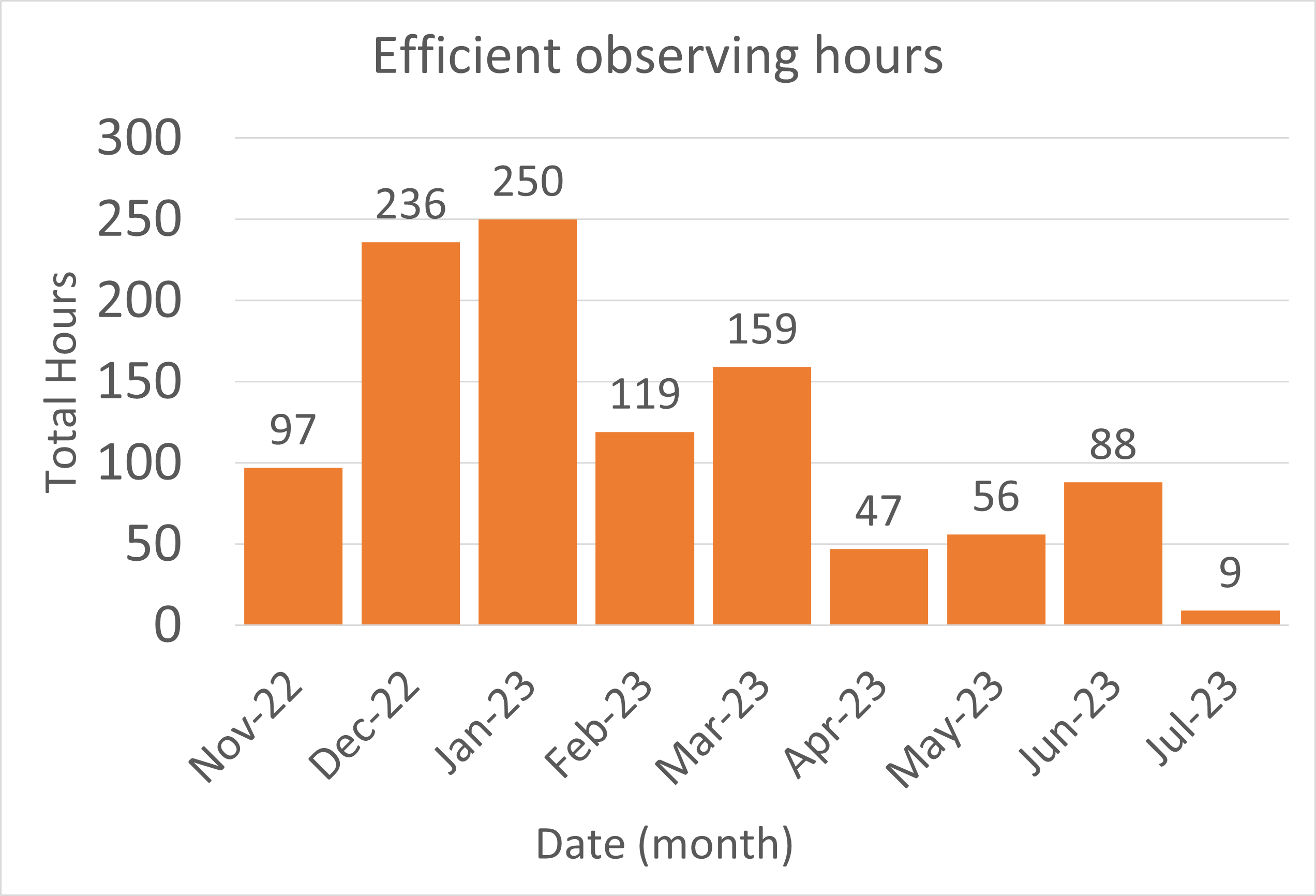}}\\
   \subfloat{
     \includegraphics[width=\textwidth]{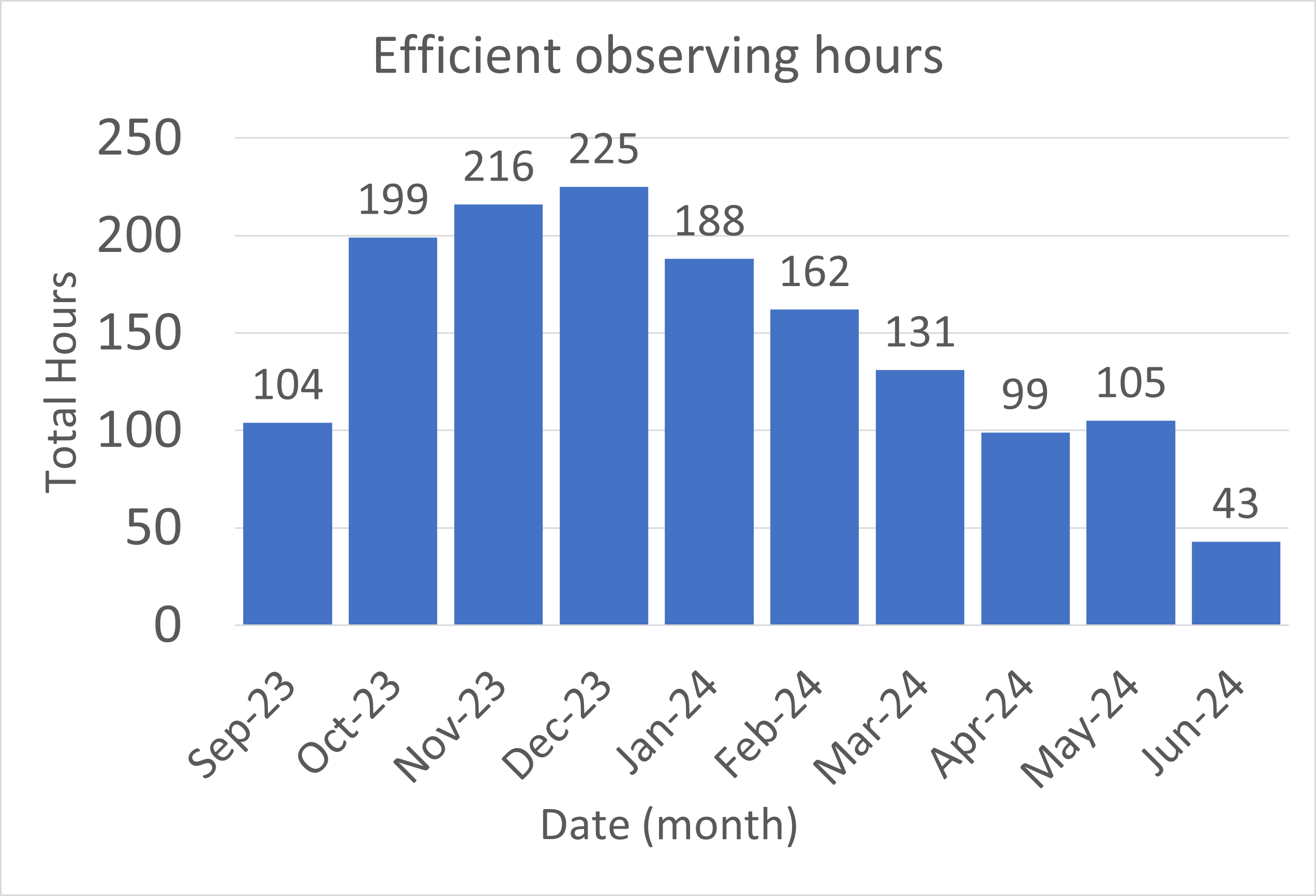}}\\
   \caption{The upper panel shows the effective observing hours from November 11, 2022, to July 1, 2023, while the lower panel displays the effective observing hours from September 1, 2023, to June 27, 2024.} 
   \label{hours}
\end{figure}

\begin{figure}
   \centering
   \subfloat{
     \includegraphics[width=\textwidth]{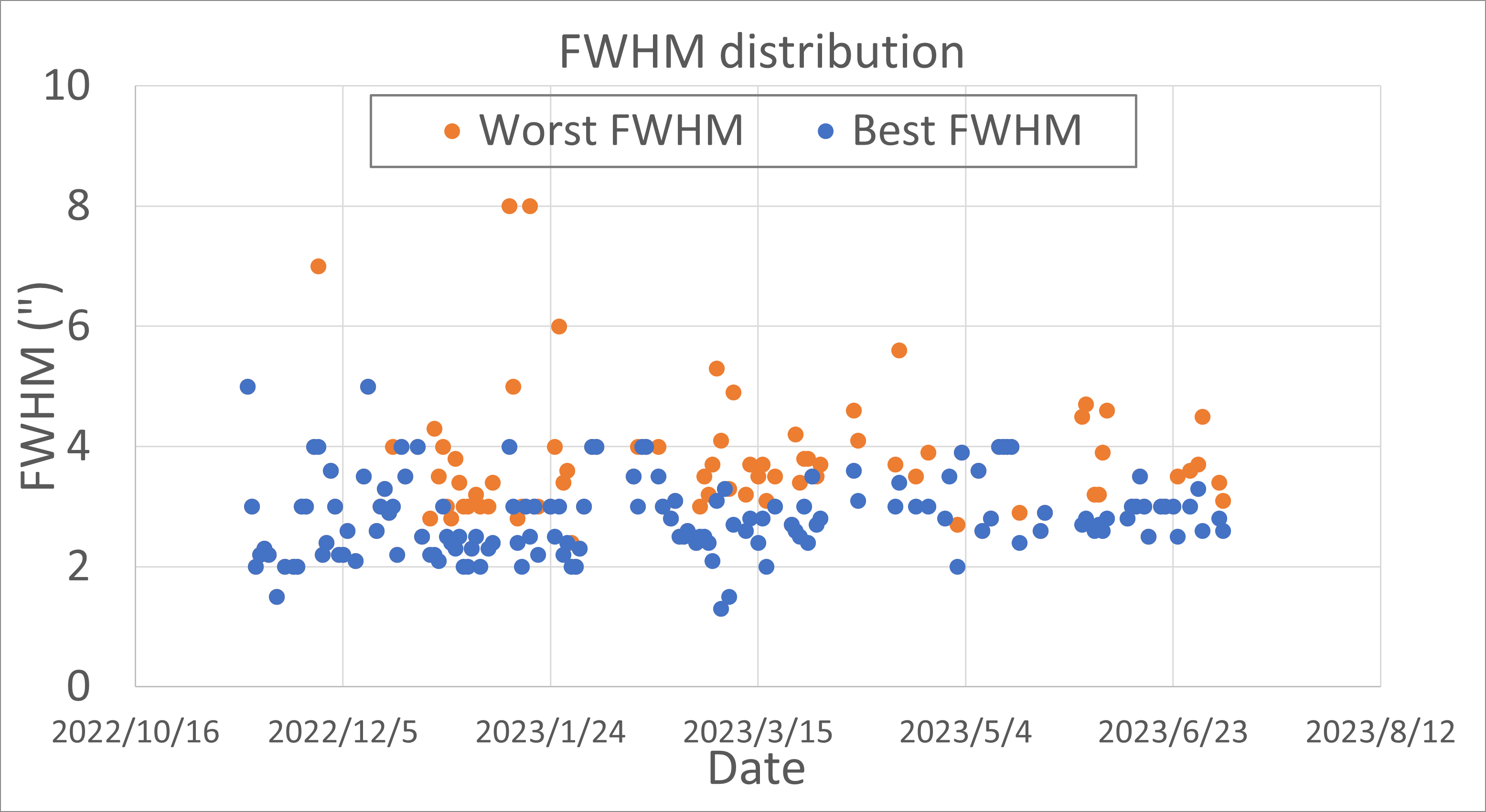}}\\
   \subfloat{
     \includegraphics[width=\textwidth]{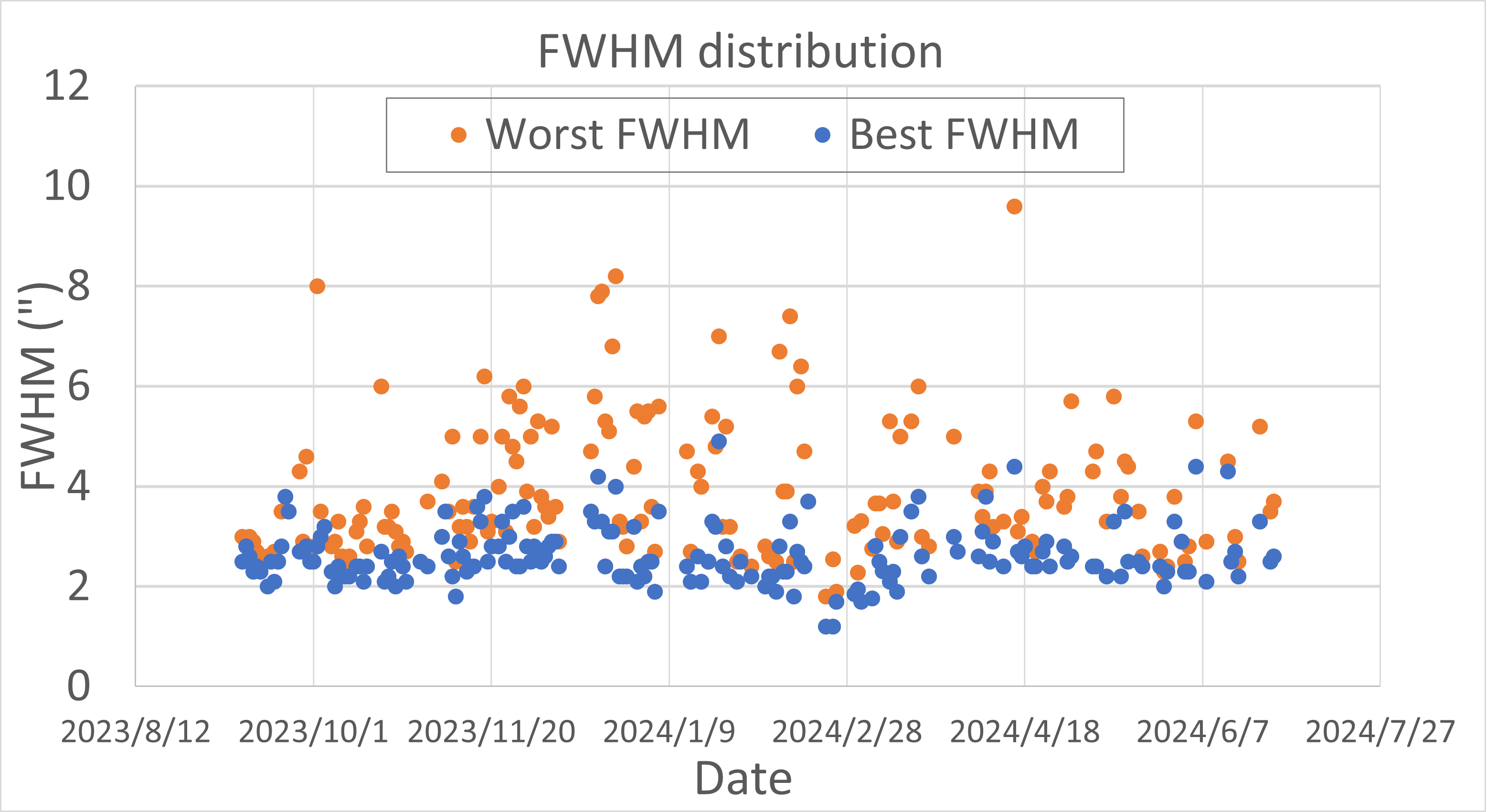}}\\
   \caption{FWHMs of the frames recorded each night by the MST. The blue dots represent the best FWHMs for each night, while the orange dots indicate the worst. The upper panel illustrates the FWHMs for the first year, and the lower panel displays the FWHMs for the second year.} 
   \label{fwhm}
\end{figure}

From November 2022 to June 2024, the MST has been in operation for 552 days, during which we have collected observational data on 344 of those days. In total, we accumulated 2,516 hours of observational data, as shown in Figure~\ref{hours}. Using the Maxim DL software, we calculated the full width at half maximum (FWHM) values for the frames captured each night, which are predominantly fall within the range of $2''$ to $3''$, as shown in Figure~\ref{fwhm}. Additionally, Figure~\ref{fwhm} also presents statistical results for the worst FWHM values recorded each night, which are mostly between $3''$ and $4''$.

\subsection{Camera tests}
\begin{figure}
  \begin{minipage}[t]{0.495\textwidth}
  \centering
   \includegraphics[width=70mm]{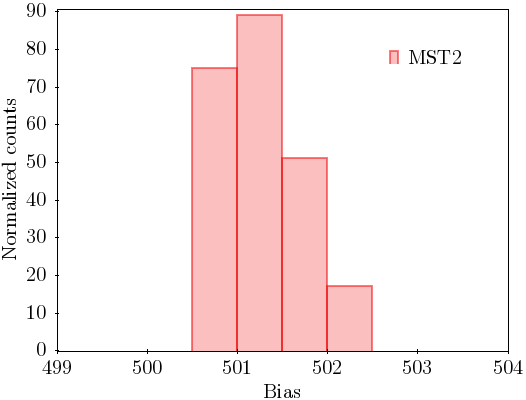}
  \end{minipage}
  \begin{minipage}[t]{0.495\textwidth}
  \centering
   \includegraphics[width=70mm]{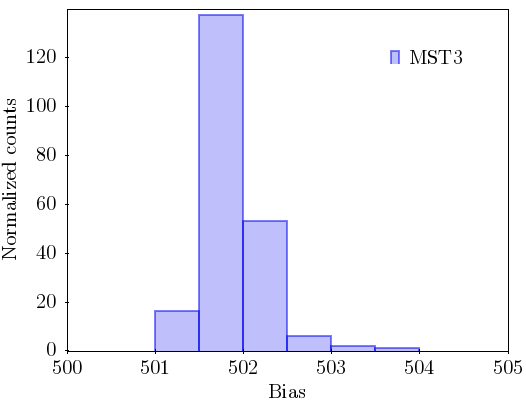}
  \end{minipage}\\
  \begin{minipage}[t]{0.495\textwidth}
  \centering
   \includegraphics[width=70mm]{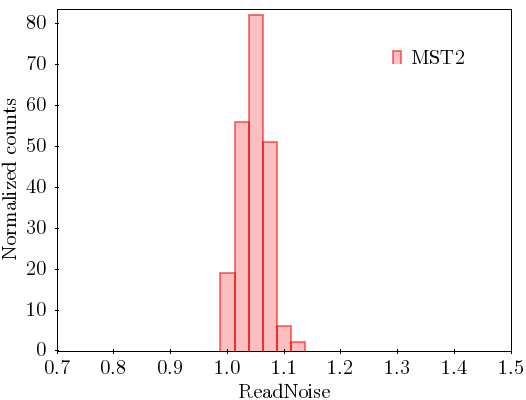}
  \end{minipage}
  \begin{minipage}[t]{0.495\textwidth}
  \centering
   \includegraphics[width=70mm]{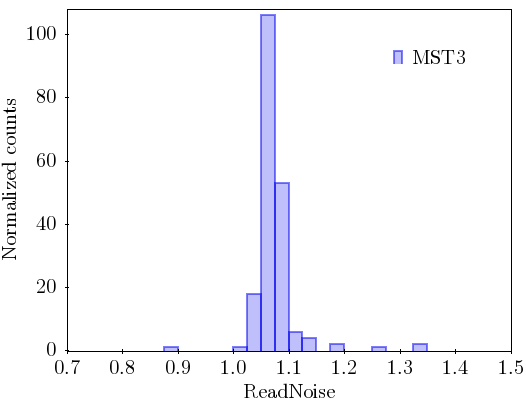}
  \end{minipage}\\
  \begin{minipage}[t]{0.495\textwidth}
  \centering
   \includegraphics[width=70mm]{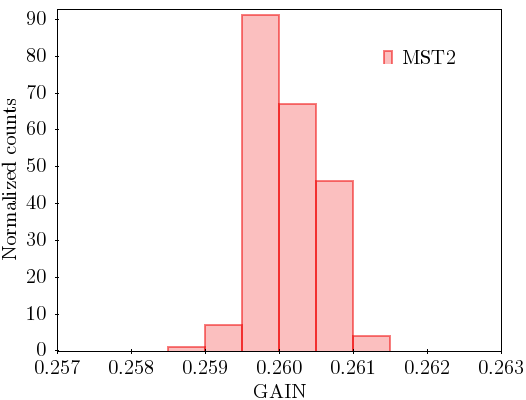}
  \end{minipage}
  \begin{minipage}[t]{0.495\textwidth}
  \centering
   \includegraphics[width=70mm]{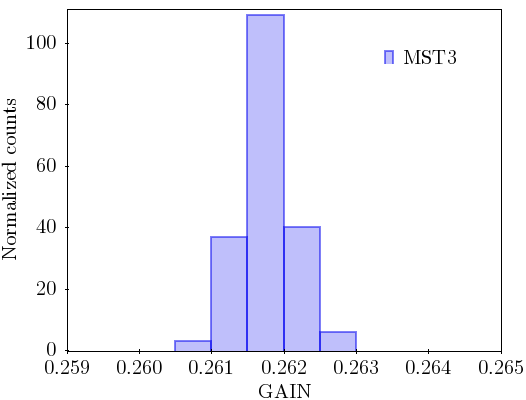}
  \end{minipage}
   \caption{The histograms of bias values, read noise values and gain values for the ZWO ASI6200MM Pro CMOS cameras on MST2 and MST3 during the first-two-year operation, indicating the stability of the two cameras.} 
   \label{bias}
\end{figure}

To assess the performance of the CMOS camera, we have conducted a series of tests, including the bias stability, abnormal response pixels, dark current, and linearity. The dark current is roughly proportional to the exposure time, $\sim$ $2\times 10^{-3}$e$^-$pixel$^{-1}$s$^{-1}$ at Temperature = $0\circ C$, which is almost negligible. For testing the linearity of the CMOS, we took a series of flat images with exposure times ranging from 0.01s to 8.1s using a stable light source, which lead to an ADU value ranging from $\sim$ 600 to saturation. When the ADU value exceeds 10,000, the accuracy of pixel values is within 0.3\%, which indicating a good linearity. A detail discussion about camera tests can be seen in \cite{zhang+et+al+2025} We have also presented statistics on the bias values, read noise, and gain of the cameras over the first-two-year operation, as shown in Figure~\ref{bias}. The results indicate that the bias values are mostly distributed around $501$, the read noise is approximately $1.06$e$^-$, and the gain is around $0.26$ e$^-$/ADU, demonstrating the stable performance of the cameras.

\subsection{Astrometry tests}
\begin{figure}
  \begin{minipage}[t]{0.495\textwidth}
  \centering
   \includegraphics[width=70mm]{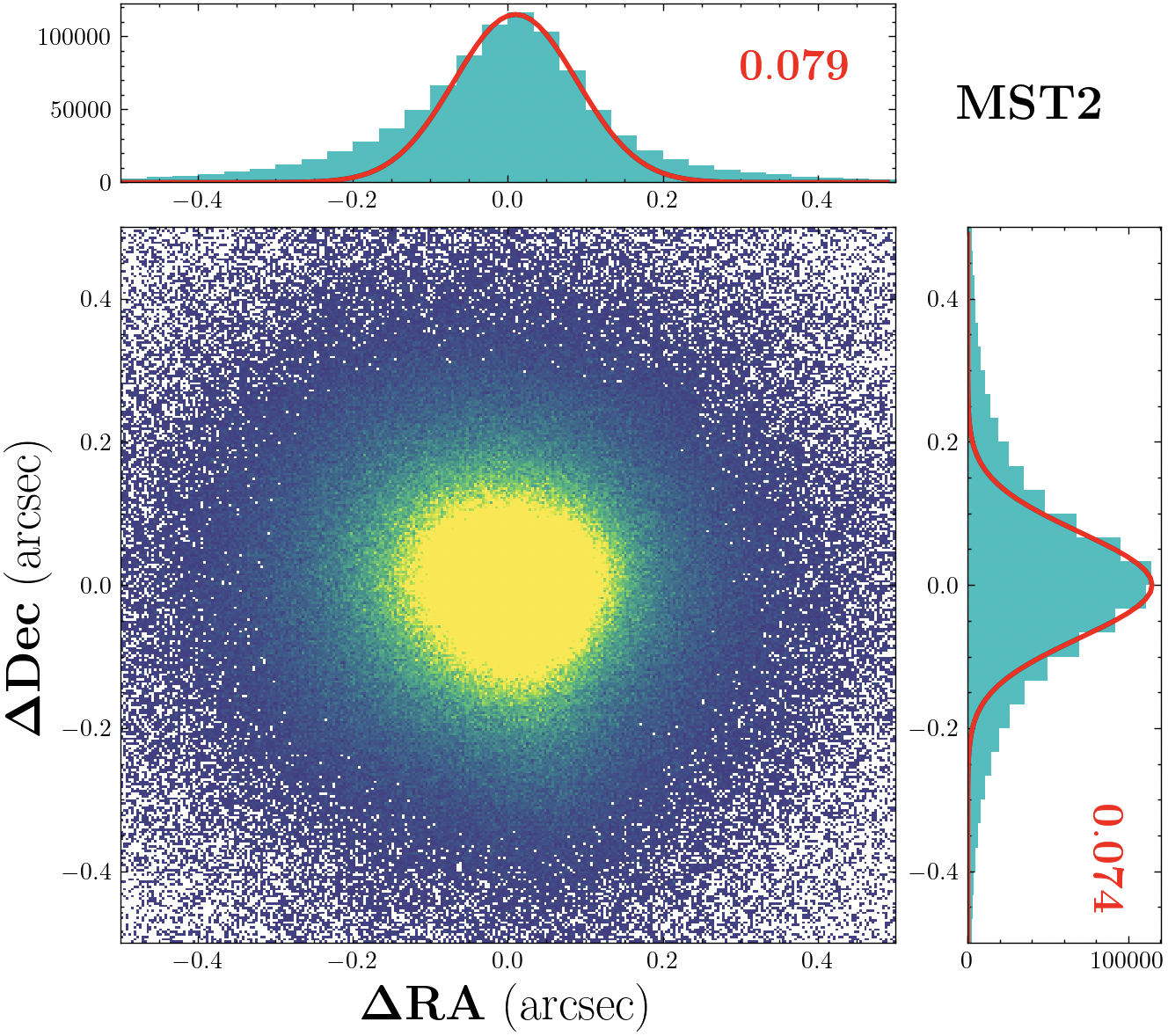}
  \end{minipage}
  \begin{minipage}[t]{0.495\textwidth}
  \centering
   \includegraphics[width=70mm]{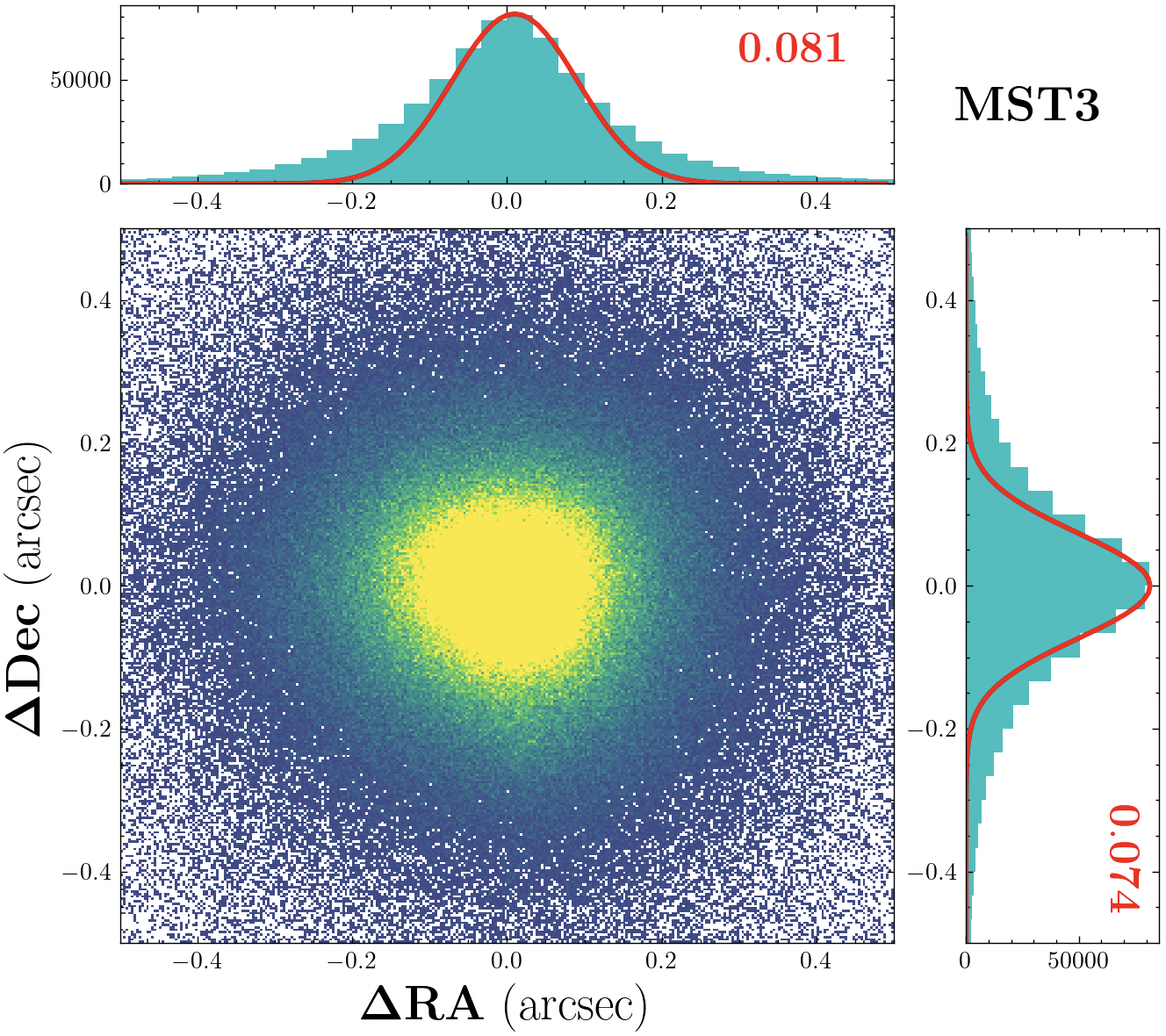}
  \end{minipage}
   \caption{Distributions of two directional position differences residuals between the MST and Gaia for the images during the first-two-year operation. MST2 and MST3 are shown in left and right panel, respectively. The position difference between the MST and Gaia DR3 remains consistent within the range of 70–80 mas.} 
   \label{stable}
\end{figure}

We also evaluated the astrometry accuracy of the MST data using Gaia DR3 by calculating the two-axis astrometric residuals for the observational data from the first-two-year operation. As shown in Figure~\ref{stable}, the position difference between the MST and Gaia DR3 remains consistent, predominantly within the range of 70–80 mas (\citealt{Xiao+et+al+2025}). 

\subsection{Photometry tests}
\begin{figure}[ht]
  \centering
   \includegraphics[width=\textwidth, angle=0]{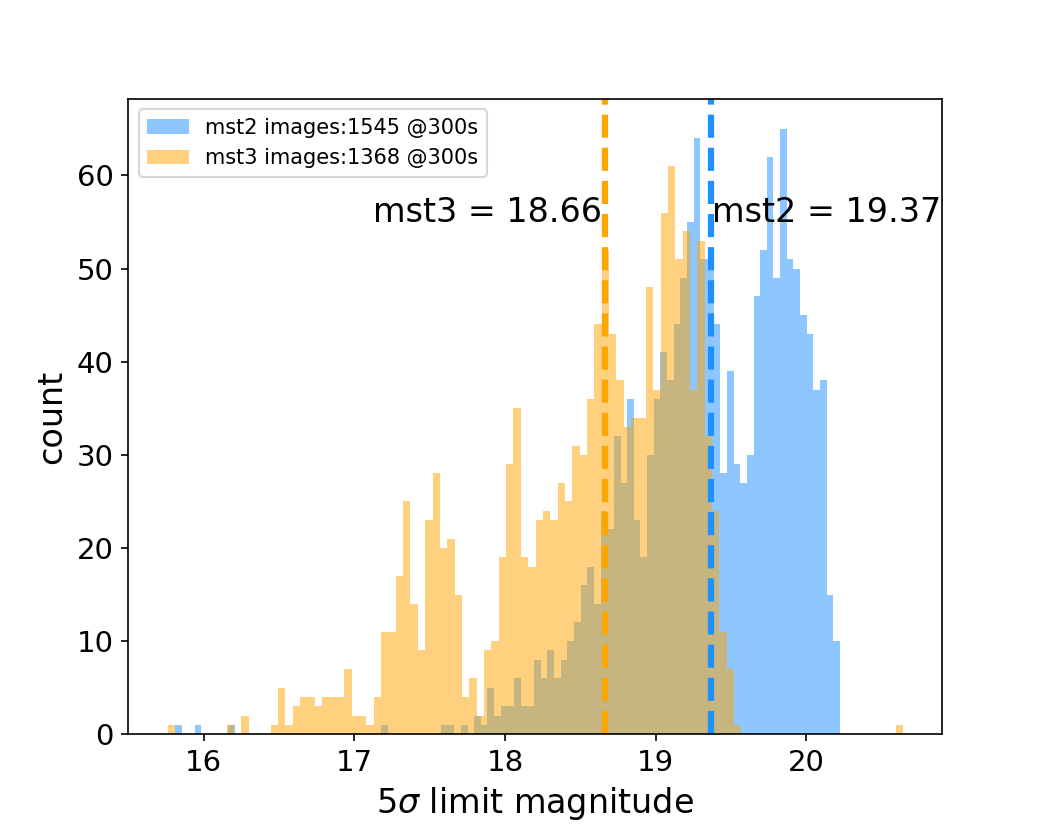}
   \caption{The histogram about limiting magnitude distributions of MST2 ($g'$-band) and MST3 ($r'$-band) during the operation.} 
   \label{limits}
\end{figure}

Figure~\ref{limits} presents the photometry of the MST, which shows that the 5-sigma magnitude limit for the images during the operation. According to the histogram in this figure, the magnitude limits can generally reach 19.37 mag in the $g'$-band and 18.66 in the $r'$-band under optimal observational condition. We have also compared the photometry results between the MST and TESS, finding consistency with TESS's results. Further details can be found in \cite{Xiao+et+al+2025}.

\section{Summary and Future plan}
\label{sect:sum}

\subsection{Summary}
\label{summary}

From November 11, 2022 to June 30, 2024, we have monitored 93 fields, and also have done the follow-up of 45 transients from TNS, 2 comets, 33 GWs alerts from LIGO. Our in-house OCS system has successfully been used during the first-two-year operation. The success of the operation has depended on the efforts and the dedication of our science, technology and observation teams. In addition to the scientific goals, in education, the MST also plays an important role for enabling students to develop skills in observation and data processing. Detailed discussions on the CMOS testing, the calibrations, the data processing and pipeline, and other science results will be presented in other published papers of this volume (\citealt{zhang+et+al+2025,Xiao+et+al+2025,Li+et+al+2025,Gu+et+al+2025,Liu+et+al+2025}).

\subsection{Future Plans}
\label{future}

The MST will continue to update the OCS system and refine observing plans to ensure thorough preparation for the future SiTian project, effectively serving as its pathfinder. Presented below are some planned tests for the MST over the next one to two years.

\subsubsection{Supernovae and TDEs survey}
Since one of the main science of SiTian project is to find transient events, rather than only follow-up these transients with other telescopes, such as ZTF. We should utilize the advantages of the SiTian project, which has rapid survey speed given its large FOV, to capture supernovae and TDEs as early as possible. We will select some sky fields with higher event rates of supernovae or TDEs for daily long-term monitoring to test the best strategy to find these types of sources (\citealt{wang+et+al+2025}).

\subsubsection{Real-time datapProcessing}
In order to capture transient as soon as possible, we need to obtain the results given by imaging subtraction between real-time image and pre-observed image templates. Therefore, rapid calibration and subtraction data processing are essential immediately after a new exposure finishes. We are developing a real-time data processing pipeline that is capable of generating preliminary results within 5 minutes of completing an exposure and presenting the outcomes as an image. When the results reveal unknown variable sources, the pipeline will automatically trigger alerts for these transients (\citealt{Gu+et+al+2025}).

\subsubsection{Auto focusing}
A rapid survey relies on stable observational quality and should not be disrupted and waste time capturing additional focusing images during operation. To address this issue, we plan to employ AI to establish a direct correlation between the point spread function (PSF) of a frame and the focuser position. Using this mechanism, the focuser position can be adjusted automatically based on the previous exposure image, eliminating the need for multiple focusing images typically required by traditional auto-focusing methods.

\subsubsection{Dynamic exposure sequence}
Currently, we still manually schedule the dynamic plan. In the future, we aim to automatically and dynamically adjust the exposure sequence based on the target's altitude over time after prioritization, and plan to integrate this module into the OCS system.

\subsubsection{MST with AI}
Since AI technology becomes more and more convenient and useful nowadays, we plan to integrate all the modules with AI and robotic systems based on a large database of scientific knowledge to reduce the reliance on human resources and make the operation significantly more efficient.

\normalem
\begin{acknowledgements}

This work has been supported by the National Key R\&D Program of China (Grant No.2023YFA1608304, No.2023YFA1608300), the Strategic Priority Research Program of the Chinese Academy of Sciences (XDB0550100, XDB0550000), and the National Natural Science Foundation of China (NSFC) grant (No. 12090041, No. 12090040, No. 11903054, No. 12422303, No. 12261141690).

The SiTian project is a next-generation, large-scale time-domain survey designed to build an array of over 60 optical telescopes, primarily located at observatory sites in China. This array will enable single-exposure observations of the entire northern hemisphere night sky with a cadence of only 30-minute, capturing true color (gri) time-series data down to about 21 mag. This project is proposed and led by the National Astronomical Observatories, Chinese Academy of Sciences (NAOC). As the pathfinder for the SiTian project, the Mini-SiTian project utilizes an array of three 30 cm telescopes to simulate a single node of the full SiTian array. The Mini-SiTian has begun its survey since November 2022. The SiTian and Mini-SiTian have been supported from the Strategic Pioneer Program of the Astronomy Large-Scale Scientific Facility, Chinese Academy of Sciences and the Science and Education Integration Funding of University of Chinese Academy of Sciences.

\end{acknowledgements}
  
\bibliographystyle{raa}
\bibliography{bibtex.bib}

   \label{lastpage}

\end{document}